\documentclass[12pt,a4paper,onecolumn,notitlepage]{article}

\newcommand*{\fref}[1]{figure~\ref{#1}}
\newcommand*{\sfref}[2]{figure~\ref{#1}(#2)}
\newcommand*{\sfrefs}[2]{figures~\ref{#1}(#2)}
\newcommand*{\latin}[1]{\textit{#1}}

\usepackage{graphicx}

\usepackage[latin1]{inputenc}

\usepackage{textcomp}
\usepackage{txfonts}

\usepackage{setspace}

\usepackage[a4paper,top=1in,bottom=1in,left=0.85in,right=0.85in]{geometry}

\usepackage[font={footnotesize},labelfont={bf,sf}]{caption}
\DeclareCaptionLabelSeparator{colon}{. }

\usepackage{notes2bib}

\usepackage{hyperref}
\hypersetup{colorlinks=true,citecolor=blue,linkcolor=blue,urlcolor=blue}

\usepackage[compress]{cite}

\usepackage{sectsty}
\sectionfont{\normalsize\sffamily}

\begin{document}

\title{\raggedright\Large\bfseries\sffamily \vspace{-1.5cm} Subcritical switching dynamics and humidity effects in nanoscale studies of domain growth in ferroelectric thin films}

\author{\normalsize\bfseries\sffamily Cédric Blaser and Patrycja Paruch \\
\small Department of Quantum Matter Physics, University of Geneva,\\
\small 24 Quai Ernest-Ansermet, 1211 Geneva 4, Switzerland \\
\small E-mail: cedric.blaser@unige.ch\\
\\
\small Published in {\it New Journal of Physics}, vol. 17, p. 013002, doi: \href{http://dx.doi.org/10.1088/1367-2630/17/1/013002}{10.1088/1367-2630/17/1/013002}\\
\small Received: 18 August 2014; Accepted for publication: 21 November 2014; Published: 9 January 2015\\
\\
\small Content from this work may be used under the terms of the \href{http://creativecommons.org/licenses/by/3.0/}{Creative Commons Attribution 3.0 licence}.\\
\small Any further distribution of this work must maintain attribution to the authors and the title of the work, \\
\small journal citation, and DOI.}

\date{}

\maketitle

\begin{spacing}{1.18}

\section*{Abstract}
Ferroelectric domain switching in \textit{c}-axis-oriented epitaxial Pb(Zr$_{0.2}$Ti$_{0.8}$)O$_3$ thin films was studied using biased scanning probe microscopy tips. While linear and logarithmic dependence of domain size on tip bias and writing time, respectively, are well known, we report an additional linear dependence on relative humidity in the 28--65\% range. We map out the switched domain size as a function of both the tip bias and the applied pulse time and describe a growth-limited regime for very short pulses and a nucleation-limited regime for very low tip bias. Using ``interrupted-switching'' measurements, we probe the nucleation regime with subcritical pulses and identify a surprisingly long relaxation time on the order of 100~ms, which we relate to ionic redistribution both on the surface and within the thin film itself.

\section*{Keywords}
ferroelectrics, polarization switching dynamics, piezoresponse force microscopy, humidity, nucleation

\section{Introduction}
\begin{figure}[!t]
\centering
\includegraphics{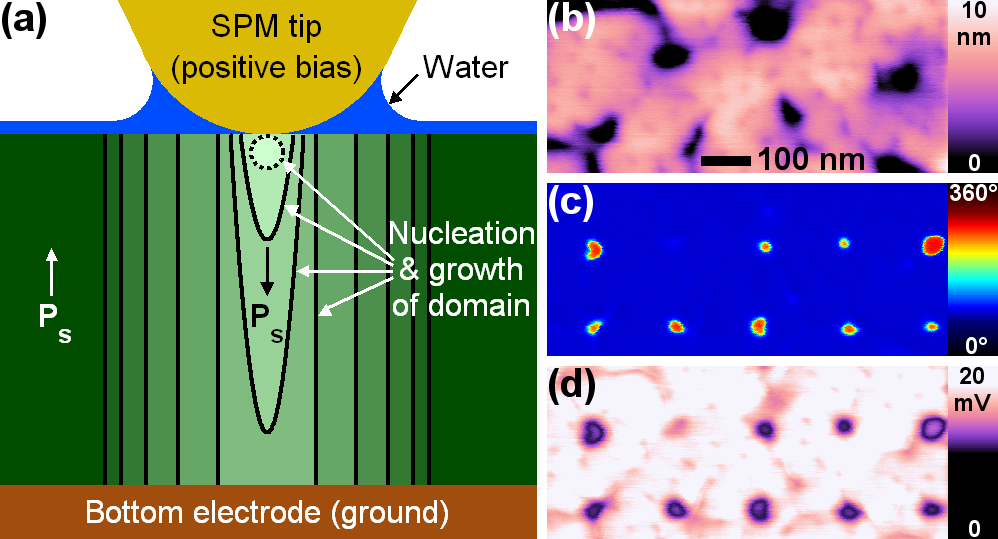}
\caption{\label{fig1}(a) Schematic of the electric field induced nucleation and subsequent growth of a domain in a ferroelectric thin film. Representative images of (b) topography, (c) PFM phase, and (d) PFM amplitude signals on our Pb(Zr$_{0.2}$Ti$_{0.8}$)O$_3$ (PZT) sample.}
\end{figure}

The characteristic electric polarization of ferroelectric materials can be locally controlled via an electric field applied with a biased scanning probe microscopy (SPM) tip \cite{paruch_apl_01_afm_arrays, cho_apl_02_AFM_Tbit}. The extremely small domains -- distinct regions with a well-defined polarization state -- formed in this fashion remain stable over extended time periods \cite{blaser_apl_12_CNT_FE}. Such nanoscale non-volatile domains allow ultrahigh (beyond 10~TBit~inch$^{-2}$) information storage densities to be envisaged \cite{tayebi_apl_10_CNT_PFM}, motivating extensive studies of their switching dynamics \cite{rodriguez_08_am_nanoscale_polarization_switching}. Thermodynamically, the switching process can be described \cite{morozovska_prb_09_nanodomain_formation, balke_natnano_09_BFO} 
as the initial, almost instantaneous, softening and reorganization of a highly localized volume of polarization under the biased SPM tip, activated by the intense electric field in this region, leading to the formation of a critical nucleus with charged domain walls, then very rapid forward propagation, and finally slower lateral expansion of the growing domain, as schematically illustrated in \sfref{fig1}{a}. Experiments have focused primarily on the latter, more accessible regimes, using piezoresponse force microscopy (PFM) to image needle-like \cite{molotskii_prl_03_domain_breakdown} or cylindrical \cite{tybell_prl_02_creep} domains in single crystals and thin films. Such studies have demonstrated the creep-like lateral motion of domain walls pinned by defects \cite{kim_afm_13_universality_switching, vasudevan_afm_13_nonlinearity, paruch_comren_13_DW_review}, recently confirmed with atomic resolution by direct imaging of domain growth using transmission electron microscopy \cite{gao_natcom_11_PZT_TEM_switching}. They also highlighted the crucial role of surface adsorbates \cite{dahan_apl_06_humidity_domains, rodriguez_prl_07_PFM_liquids, shur_jap_11_adsorbates_LNO, ievlev_apl_14_humidity_LNO}, particularly water, which determine the electrostatic boundary conditions and thus the size, shape and growth rate of the domain \cite{maksymovych_prl_09_nucleation_disorder, ievlev_natphys_14_domain_interactions, guyonnet_environmental}. In contrast, the subcritical regime, where the voltage pulses are too short to fully stabilize domains of inverted polarization, is much less well understood, with the additional complexity of highly dynamic electrochemical effects such as charge injection or re-ordering under the biased SPM tip \cite{kalinin_nano_11_electrochemical_SPM}. These different processes all take place in a very confined region of high electric field, at possibly very different time scales, and can strongly influence the initial stages of polarization softening and stabilization of the critical nucleus.

In this letter, we present a detailed PFM mapping of domain switching in epitaxially-grown Pb(Zr$_{0.2}$Ti$_{0.8}$)O$_3$ (PZT) as a function of tip bias, writing time, and relative humidity in the range from 28\% to 65\%. For low tip bias, we find that successful switching occurs only when the activation threshold for polarization reversal is itself comparably low, and thus depends very sensitively on the local, highly heterogeneous defect distribution \cite{jesse_natmat_08_SSPFM, Kalinin_AM_10_defect_mediated_switching} which provides a randomly varying pinning/activation energy landscape \cite{paruch_comren_13_DW_review}. However, when these activation threshold requirements are met, and as long as the tip bias is maintained, the resulting domains nonetheless grow outward by radial domain wall motion, up to significant size. For short writing times and high tip bias, in contrast, we observe a very high probability of initial nucleation, but successful switching depends on the stabilization of this critical nucleus under the SPM tip. The resulting domains are very small. We refer to these two regimes as nucleation-limited and growth-limited, respectively, in analogy to switching behavior observed in ferroelectric capacitors \cite{Sharma_Nanotech_13_switching_asymmetry} and thin films \cite{guyonnet_am_11_DW_conduction}. Finally, using very short repeated pulses well below switching limits, separated by a defined ``interruption time'', we probe the subcritical regime of domain switching, demonstrating that the initial polarization softening and subcritical nucleus formation are characterized by surprisingly long relaxation times of over 100~ms.

We used a 270~nm thick epitaxial, $c$-axis oriented ferroelectric PZT thin film grown by off-axis radio-frequency magnetron sputtering on 35~nm thick conducting SrRuO$_3$ on (001) single-crystal SrTiO$_3$ (\textit{CrystTec}), with high crystalline quality \cite{gariglio_apl_07_PZT_highTc} and 2.4~nm root-mean-square surface roughness. The remanent polarization of 75~\textmu{}C~cm$^{-2}$ is perpendicular to the film plane, and monodomain (`up-polarized') as grown, showing an imprint of $-1.6$~V with coercive voltages of +3.3~V and $-6.5$~V at a cycle frequency of 500~Hz. To switch the polarization, voltage pulses were applied to \textit{Bruker MESP} tips with the SrRuO$_3$ layer grounded, using a \textit{Veeco Dimension~V} atomic force microscope \bibnote{The pulse time accuracy of the \textit{NanoScope~V} controller was monitored using a \textit{Tektronix TDS~2014B} oscilloscope and software corrected to 0.1~\textmu{}s.}. PFM imaging with simultaneous recording of topography (\sfref{fig1}{b}), phase (\sfref{fig1}{c}) and amplitude (\sfref{fig1}{d}) signals was performed in contact resonance using a drive frequency 10~kHz above the resonance peak ($\sim$300~kHz) at a drive amplitude of 2.0~V. The effective domain radius was extracted from the total area obtained from the binarized PFM phase images. A home-made casing around the microscope, combined with either desiccant material (silica gel beads) or evaporative water source allowed the relative humidity (RH) of the sample environment to be stabilized in the range between 28\% and 65\%. The ambient relative humidity of the laboratory was measured to vary between 30\% and 54\% during the course of one year.

\section{Results and Discussion}

\begin{figure*}[!ht]
\centering
\includegraphics{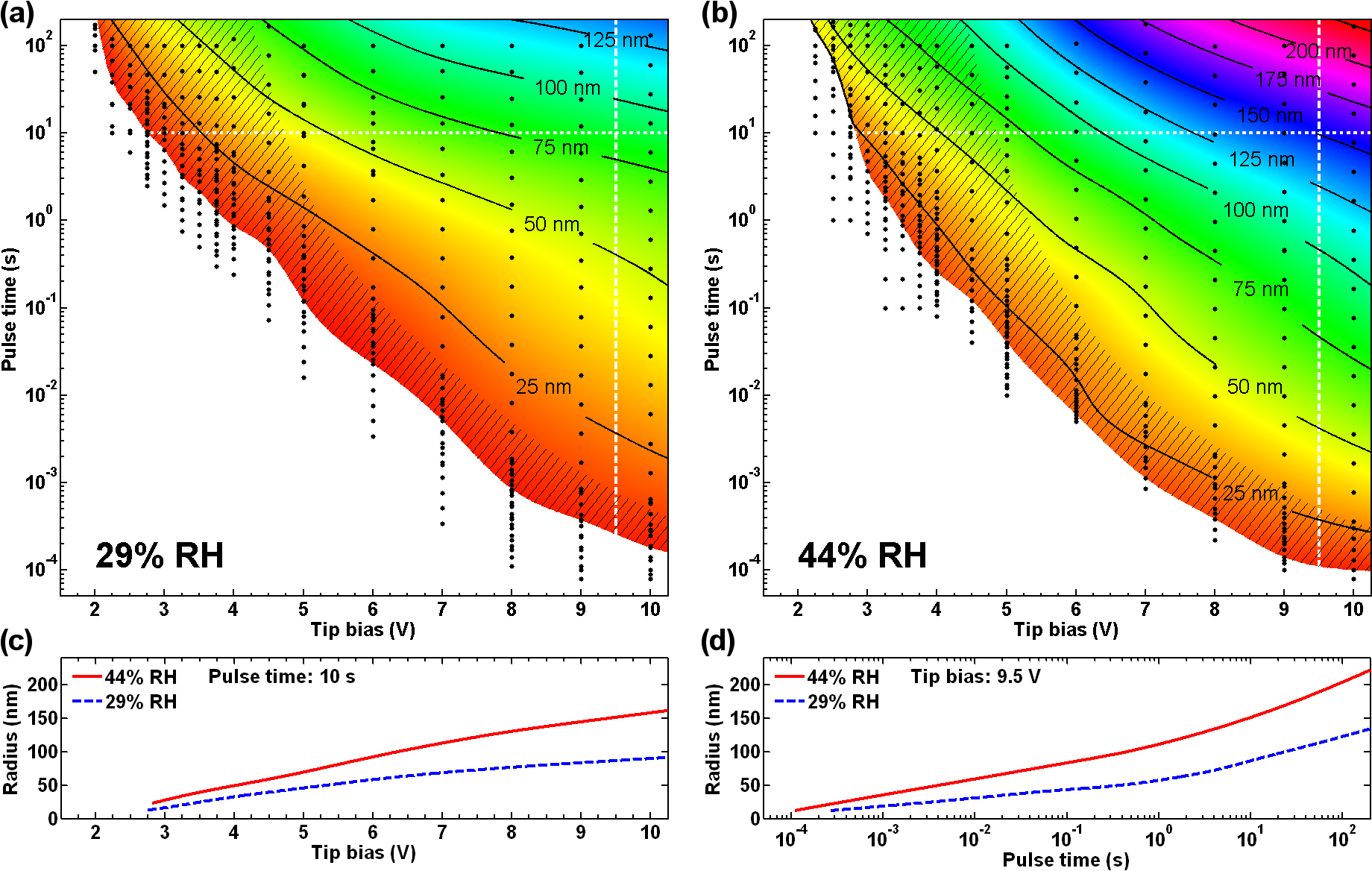}
\caption{\label{fig2}Average effective domain radius as a function of tip bias and pulse time obtained at a relative humidity of (a) 29\% and (b) 44\%. Iso-size lines and color shading indicate the parameters for which domains have the same radius. Hatching represents switching success rates below 100\%. No switching was observed in the white region. Black dots represent each measured set of parameters. (c) Cross-sections along the white horizontal dotted lines through the parameter space showing the linear dependence of domain radius on tip bias. (d) Cross-sections along the white vertical dashed lines through the parameter space showing the logarithmic dependence of domain radius on pulse time.}
\end{figure*}

To map out the switched domain radius as a function of the writing parameters, we applied a tip bias between 2 and 10~V for writing times ranging from 80~\textmu{}s to 600~s. For each set of parameters measured, represented by a black dot in \fref{fig2}, five separate switching attempts were made, and the size of the resulting successfully switched domains averaged to obtain the effective radius. Successful switching was defined as a 180° PFM phase contrast between the two polarization states, and the presence of a well-defined ring feature corresponding to a decrease of PFM amplitude at the domain wall, characteristic of a columnar domain penetrating through the full film thickness. At both RH values, we recognize the previously-reported \cite{paruch_comren_13_DW_review} linear (\sfref{fig2}{c}) and logarithmic (\sfref{fig2}{d}) dependence of domain size on tip bias and writing time, respectively. However, for comparable choice of writing parameters, domain radii at 44\%~RH (\sfref{fig2}{b}) are almost twice as big as those obtained at 29\%~RH (\sfref{fig2}{a}). The minimum stable domain radius, almost independent of the tip bias for an appropriate choice of pulse time, is around 17~nm for 44\%~RH and only 10~nm for 29\%~RH. These very small domains remained stable for at least 6~weeks. The hatched area represents the writing parameters for which the success rate of domain switching is less than 100\%. In this region of parameter space, comparable for both high and low RH, there is a progressive transition from certainty of successful switching, with the resulting domain penetrating through the film thickness, to a regime where the switching and stabilization of a domain become rare. We note that unsuccessful switching includes observations of no change in either phase or amplitude in subsequent PFM imaging, and of trace features, such as can be seen in the upper row of domains in \sfref{fig1}{c, d}, which nonetheless fail to meet the full criteria for successful switching.

For high tip bias but short writing time, the latter appears to be the key limiting factor. In this region of parameter space, partially switched needle-like domains \bibnote{See Supporting Information for detailed definition of partially and successfully switched domains.}, with charged domain walls, which do not penetrate through the full sample thickness can often be detected, disappearing between hours and a few days after writing. Such partially switched domains are characterized by an intermediate PFM phase contrast of less than 180° and a lower PFM amplitude signal without a distinct ring feature. Their presence demonstrates that here stabilization and subsequent growth of the critical nucleus, rather than the initial nucleation step, limit domain switching. For low tip bias and long writing time, in contrast, we detect no trace of partial switching. Rather, any domains observed are fully stabilized, sometimes with diameters well above 100~nm, although the probability of their formation gradually approaches 0\%. The limiting factor in this region of parameter space is thus the low tip bias, insufficient to meet the threshold requirement for nucleation. This nucleation threshold, shown to locally vary by over 6~V in epitaxial PZT \cite{jesse_natmat_08_SSPFM}, is determined by the distribution of defects in the thin film. The domain switching probability thus depends very sensitively on the local defect landscape. We note that the upper limit of the hatched area does not follow the iso-size lines of equal domain radius, thus for a tip bias below 4.5~V, domain switching is not certain even at 100~s pulse times.

\begin{figure}[!ht]
\centering
\includegraphics{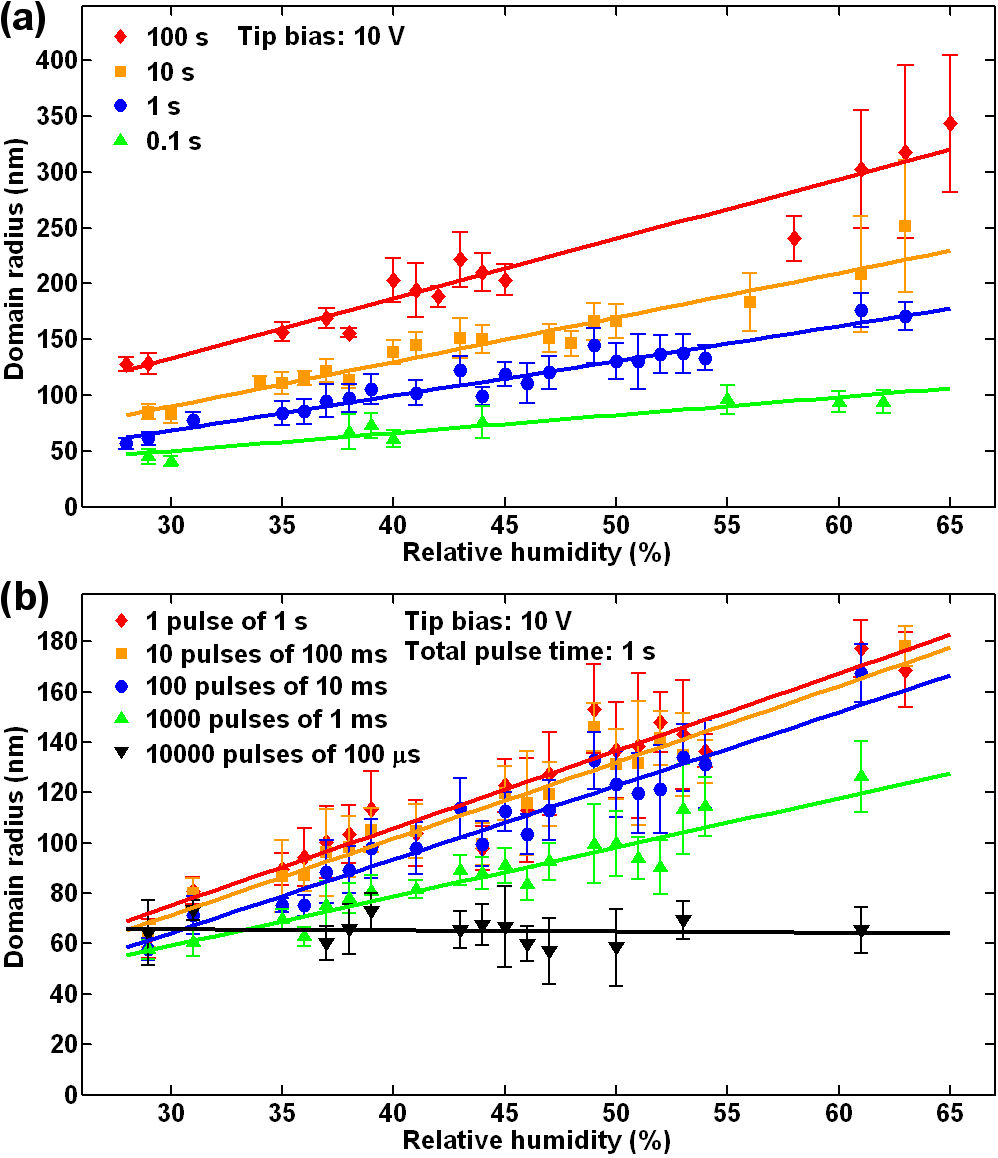}
\caption{\label{fig3}(a) Linear relation between the domain radius and the relative humidity for pulses of 0.1, 1, 10, or 100~s, at a tip bias of 10~V. (b) Domain radius as a function of relative humidity for pulses with a total time of 1~s at a tip bias of 10~V. Pulses were applied singly, or split into 10, 100, 1000, or 10\,000 intervals with an interruption time of 60~ms.}
\end{figure}

To better understand the specific contributions of the different parameters, including humidity, we applied pulses of 0.1, 1, 10, or 100~s with a tip bias of 10~V to the sample, and measured the effective radius of over 1500 domains across the available RH range \bibnote{We also evaluated the effects of progressive tip wear and variations in writing protocol of lifting, scanning, or keeping the grounded probe tip stationary after bias application. However, as detailed in the Supporting Information, these variables do not appear to play a significant role.}. As shown in \sfref{fig3}{a}, we find a linear increase of domain size with the relative humidity in the 28--65\% range, with domain radii at high RH being more than twice as big as those at low humidity. For different pulse times the slope of this linear dependence varies from 1.6 to 5.3~nm~\%RH$^{-1}$. Since the vast majority of PFM measurements are carried out at ambient conditions, this very significant effect should be recognized in any quantitative study of domain switching and growth, especially given the annual RH variation in a standard laboratory. We note that a simple extrapolation of the observed linear dependence of domain size on humidity would yield zero-size switching limits in the 5--8\%~RH range. However, domain switching has been successfully demonstrated even under ultrahigh vacuum \cite{maksymovych_sci_09_TER_PZT, guyonnet_environmental}. We therefore expect a deviation from the linear behavior for lower humidities approaching 0\%~RH. For long writing times at high humidity, we also observe an increasing spread in the measured domain size. Radii of domains written with 100~s pulses below 30\%~RH present a standard deviation of only 3--10\%, while for domains written above 60\%~RH a much greater standard deviation of up to 25\% is extracted. This increasing spread suggests a greater influence of local variations in the defect landscape, which would be expected as the electric field intensity and thus domain wall speed decreases further away from the tip. 

A straightforward explanation of the influence of the relative humidity on the growth of ferroelectric domains is readily provided by the presence of a water meniscus at the point of contact between the SPM tip and the sample, as imaged by Weeks \latin{et al.} \cite{weeks_langmuir_05_meniscus_shape_AFM_tip}. However, the several-100-nm-wide menisci reported by these authors were formed only after RH saturation, then progressive decrease to a lower target value. As long as the system stayed in a RH range below 60\%, corresponding to standard ambient conditions, the size of the meniscus at the SPM tip in fact remained below the 50~nm imaging resolution of the experiment. Thus, a theoretical estimate of the meniscus size obtained using the Kelvin equation for capillary condensation \cite{butt_meniscus_shape} provides a more useful approach. In our case, with the assumption of a 1~nm thick adsorbed layer of surface water on top of the sample \cite{asay_surface_layer}, we calculate the horizontal extent of the meniscus to range from 12 to 16~nm for RH between 28\% and 65\% \bibnote[meniscus_field]{See Supporting Information for details of water meniscus calculations and electric field simulations.}. In electrostatic simulations using finite element analysis with \textit{\mbox{COMSOL} Multiphysics} its presence indeed both increases the lateral extent over which high values of the vertical electric field (along the out-of-plane polarization axis) are maintained, and decreases the maximum field intensity at the tip-sample contact point. A water meniscus would thus lower the probability of nucleation, which depends on the intensity of the electric field in the local region under the biased SPM tip, but would promote more rapid domain growth through the high field region were nucleation to occur, in qualitative agreement with the experimental results. However, the calculated effect remains very local, with no significant change in the field profile at distances beyond 35~nm from the contact point \bibnotemark[meniscus_field]. In contrast, experimentally we observe that domain growth is still linearly influenced by humidity up to distances corresponding to the largest measured domain radii of 350~nm, suggesting that the meniscus size is only a part of the physical mechanism and that longer range effects need to be considered.

One possibility proposed by Brugère \latin{et al.} \cite{brugere_jap_11_domain_growth_simulation} is that the electric field itself slowly propagates through the weakly conductive surface water layer, driving lateral domain growth. However, Guyonnet \latin{et al.} \cite{guyonnet_environmental} demonstrated qualitatively similar domain wall creep both in ambient conditions and under ultrahigh vacuum, disfavoring this hypothesis. Their work highlighted instead the important screening effects of surface water with respect to both the ferroelectric dipolar forces, and to defects which can act as nucleation and pinning sites. Varying the strength of this screening in a simple Ginzburg-Landau switching model with numerically introduced disorder reproduced the order-of-magnitude difference in the size of the resulting domain with identical tip bias and writing voltage. We also note that under the locally very intense electric field of the biased SPM tip, the disorder potential itself can dynamically evolve. Electrochemical effects such as oxygen vacancy injection or (re)ordering, surface charging, and even irreversible physical damage to the sample \cite{kalinin_nano_11_electrochemical_SPM} can significantly change the defect landscape during switching, particularly at ambient conditions. The resulting complex interplay between surface water screening and defect dynamics would influence not only the lateral growth of domains, but also the initial stabilization of the critical nucleus.

To further explore these effects we used ``interrupted-switching'' PFM, measuring domain growth after the application of 10 V bias to the SPM tip for a total time of 1~s, subdivided into 10, 100, 1000, or 10\,000 shorter pulses separated from each other by an interruption time of 60~ms, during which the tip was grounded. Yang \latin{et al.} \cite{yang_apl_08_split_pulses} reported that a similar use of pulse splitting into shorter intervals had no effects on the growth of the resulting domain. However, as the constituent pulses approach the limit for switching, on the order of 10--100~\textmu{}s in the case of our sample, we would in fact expect significantly reduced domain sizes and lower probability of successful switching. Such short pulses should thus allow us to probe the initial stabilization of the critical nucleus under different RH.

As can be seen in \sfref{fig3}{b}, subdivision of the total pulse time into intervals longer than 10~ms indeed has relatively little effect. While the resulting domain size may be slightly smaller than that written with a single pulse, the linear dependence of domain size on RH is maintained across the measurement range. However, we find that the role played by RH becomes less important when the total pulse time is subdivided into intervals shorter than 1~ms. In fact, for 10\,000 constituent pulses of 100~\textmu{}s the final domain size of 65--70~nm appears to be independent of the humidity. We confirmed these observations in repeated experiments for total pulse times of 10 and 100~s, and also investigated the effect of varying interruption times (60, 120, 280, and 560~ms) which produced no significant change in final domain radius \bibnote[protocols]{See Supporting Information for measurements with different total pulse time, tip bias, or writing protocol.}. In all these measurements with 10~V tip bias, the probability of successful switching was almost 100\%, even for the shortest subdivision times. In contrast, we find that lowering the tip bias very strongly decreases this probability. In identical interrupted-switching PFM measurements with 8 and 6~V tip bias, we again observe that the subdivision of the total pulse time into many very short intervals results in decreased domain size, with little or no RH dependence. However, the probability of successful domain switching after 10\,000 pulses of 100~\textmu{}s is only 45\% at 8~V, and the value drops to 15\% for 6~V. For the lower (6~V) tip bias, domain writing using fractional pulses is also much more sensitive to tip wear and higher humidity levels \bibnotemark[protocols].

\begin{figure}[!ht]
\centering
\includegraphics[scale=1.5]{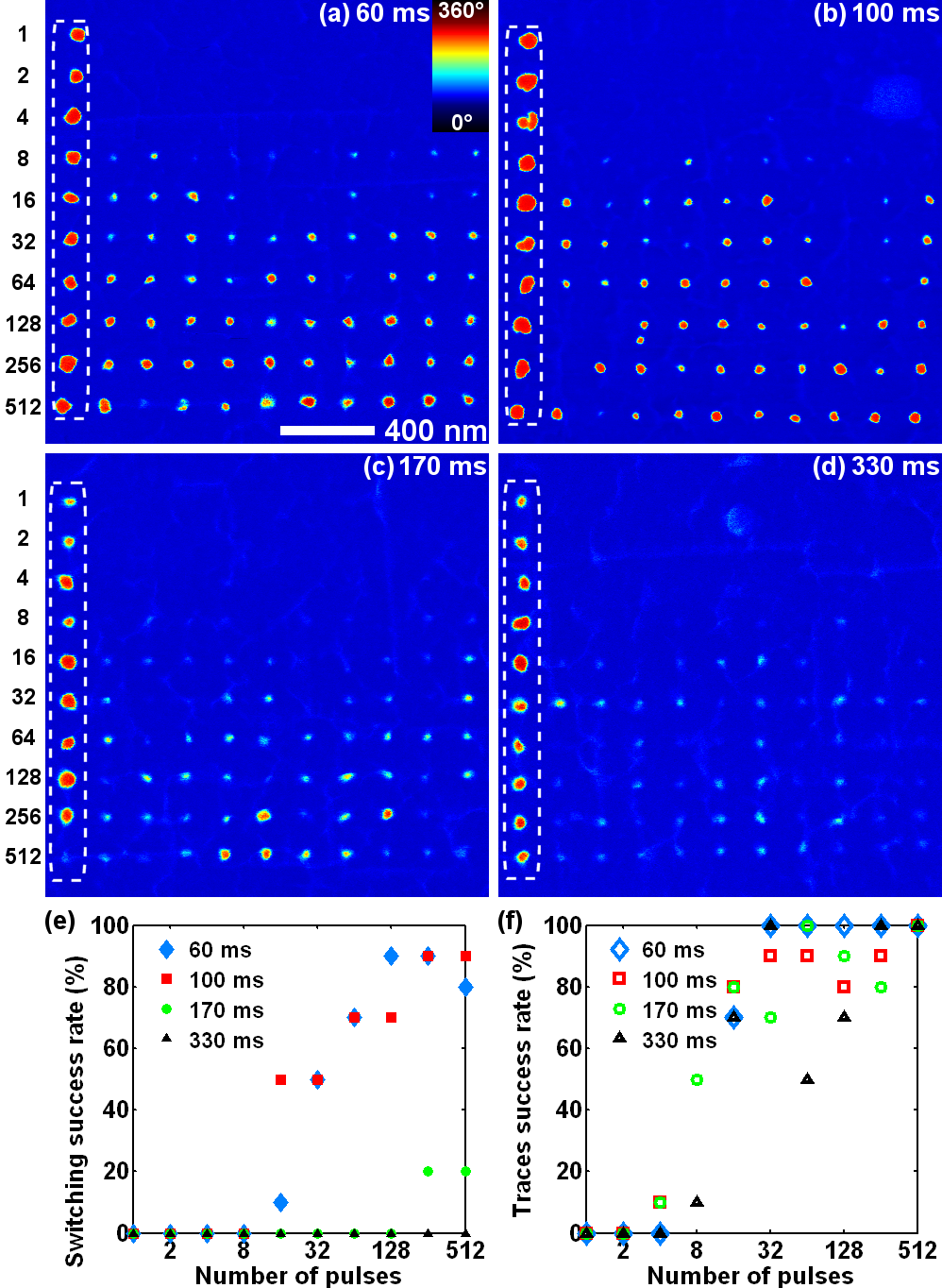}
\caption{\label{fig4}PFM phase images of domains written with multiple pulses of 10~\textmu{}s at 10~V tip bias at 33\%~RH. The number of individual pulses (indicated at left) ranges from 1 to 512, and the interruption time is either (a) 60~ms, (b) 100~ms, (c) 170~ms, or (d) 330~ms. The leftmost column of each array (framed by dashed box) was written with single 10~ms pulses and serves as marker to locate the lines. Success rates at which (e) complete domain switching and (f) traces of partial switching can be seen in the phase signal as a function of the number of individual pulses, for the above mentioned interruption times.}
\end{figure}

These results, in line with the data presented in \fref{fig2}, confirm that nucleation and full stabilization of a narrow columnar domain under the biased SPM tip occur at relatively short time scales on the order of 100~\textmu{}s. However, for the subsequent lateral growth and stabilization of large domains, we find that both high RH and longer continuous exposure to the electric field are needed. In addition, for similar pulse times and lower tip bias the resulting written domains are not only smaller, but also characterized by decreased switching probability, pointing to the strong influence of local variations of the defect landscape. 

Finally, to access the subcritical regime of the switching dynamics we used interrupted-switching PFM with constituent pulses below the switching limit, with varying interruption times. Each individual 10~V pulse was only 10~\textmu{}s long, 10 times shorter than the threshold pulse duration necessary to successfully switch a domain. As shown in \fref{fig4}, in each case we wrote a domain array, doubling the number of single pulses at each line, starting at 1 (top) and ending at 512 (bottom). The interruption times ranged from 60~ms (\sfref{fig4}{a}) to 330~ms (\sfref{fig4}{d}). For single 10~\textmu{}s, 10~V pulses we detect no visible effect on the ferroelectric film during subsequent imaging in either PFM phase or amplitude. However, with a repeated application of as few as 16 such pulses separated by interruption times of up to 100~ms, successful domain switching can be observed, presenting the minimum PFM amplitude ring feature characteristic of well-defined domain walls \bibnote{See Supporting Information for the PFM amplitude images.}. Moreover, even down to 4 repeated pulses, we can observe traces of partial switching, with intermediate PFM phase contrast and a somewhat amorphous PFM amplitude drop at the position where the bias pulses were applied. As presented in \sfrefs{fig4}{e,~f}, longer interruption times decrease both the complete and partial switching success rates, determined by counting how many domains out of the 10 in each line were fully stabilized or partially switched. For interruption times of 170~ms, repeated application of 256 and 512 voltage pulses still allowed successful domain switching, but no stable domains were observed with 330~ms interruption times. Nonetheless, even with these very long interruption times, we still detect a high rate of partial switching.

The cumulative effect of individual pulses, which in themselves are too short to have any measurable effect, proves that the as-grown state of the thin film is nonetheless modified after exposure to the electric field of the SPM tip for 10~\textmu{}s, rendering it sensitive to further excitation. These highly local changes associated with the formation and stabilization of the critical nucleus in the first stage of switching persist for surprisingly extended periods. To fully stabilize a columnar domain through the film thickness, the minimum total pulse time (160~\textmu{}s) is comparable to the 100~\textmu{}s required with a single continuous pulse, suggesting that the changes induced by each constituent pulse are conserved even with interruptions which are 10\,000 time longer than the constituent pulses themselves.  As the interruption time is increased to 170~ms, however, the required total pulse time for the same result is 25 times longer, showing that in this case the film returned at least partially towards its initial state during the interruptions. We thus estimate the relaxation time to be on the order of 100~ms. 

This relaxation time is far beyond what would be expected for either electronic or even lattice accommodation, or for rapid ``back-switching'' observed as domain walls rapidly adjust their position when the local electric field of the tip is switched off \cite{gao_natcom_11_PZT_TEM_switching}, and points to slower dynamic effects. In addition, the formation and initial stabilization of the subcritical nucleus appears to be largely RH independent within the measured range. One possibility is very local displacements of H$^+$ and OH$^-$ species in the immediate neighborhood of the tip \cite{ievlev_natphys_14_domain_interactions, strelcov_am_14_surface_charge}, where the specific shape and extent of the water meniscus plays no significant role. These charged species, formed at even the lowest partial pressures of water vapor on ferroelectric surfaces \cite{shin_nl_09_adsorbates_switching}, have been previously shown to play a critical role in screening and polarization switching kinetics \cite{fong_prl_06_PTO_chem}. Another key factor is the redistribution of oxygen vacancies, for which drift velocities of 5~pm~s$^{-1}$ under fields on the order of MV~cm$^{-1}$ have been reported in PZT capacitors \cite{gottschalk_jap_08_ovac_kinetics}. These defects, common in all perovskite oxides, are particularly abundant in thin films grown on SrTiO$_3$ substrates \cite{yuan_apl_09_Ovac_distribution}, and tend to reorganize under the influence of a biased SPM tip \cite{kalinin_nano_11_electrochemical_SPM}. Even small changes in the oxygen content in perovskite materials have been recently shown to have significant effects \cite{Xie_oxygen_vac, Biegalski_APL_2014} in terms of resistivity or c-axis parameter. A localized increase in the density of oxygen vacancies under the tip could thus strongly contribute to the observed long-life ``activation'' of this region and its subsequent tendency to switch more easily.

We also note that recent measurements of the role of RH in switching on thinned LiNbO$_3$ single crystals, generally containing fewer defects, showed completely opposite behavior with no observed change in domain size for 0--60\%~RH and a precipitous drop thereafter \cite{ievlev_apl_14_humidity_LNO}. These differences strongly suggest that electrochemical contributions from within the ferroelectric, as well as ionic effects of adsorbates on the surface, need to be considered.

\section{Conclusions}
From a detailed mapping of the size of SPM-tip-written ferroelectric domains in PZT as a function of both tip bias and pulse time for which it is applied, we identify nucleation- and growth-limited switching regimes at low bias and short writing times, respectively. We show that these regimes depend on very different factors, the first controlled by variations in the defect landscape which locally modify the nucleation bias threshold, the second by the dynamics of the stabilization of the critical nucleus and the initial stages of vertical domain growth through the film thickness. In addition, we quantify the linear dependence of domain size on the relative humidity over and beyond the full range of values corresponding to its annual variation in a standard laboratory, which can significantly affect the measurements of domain switching dynamics in ambient conditions. Finally, introducing interrupted-switching PFM in the subcritical regime, we extract surprisingly long relaxation times on the order of 100~ms for the local modifications in the ferroelectric film associated with the formation and stabilization of the critical nucleus.

\section*{Supporting Information}
Effects of tip wear, geometry of the water meniscus, finite element simulations of the vertical electric field, alternative interrupted-switching PFM protocols, and PFM amplitude images of the subcritical switching experiments. This material is available free of charge via the Internet at \href{http://stacks.iop.org/njp/17/013002/mmedia}{stacks.iop.org/njp/17/013002/mmedia}.

\section*{Acknowledgements}
The authors thank S.~Gariglio for the PZT samples and M.~Lopes, S.~Muller, and I.~Gaponenko for technical support. This work was funded by the Swiss National Science Foundation through the NCCR MaNEP and Division II grant No. 200020-138198.

\end{spacing}

\begin{spacing}{0.2}
\bibliographystyle{new_journal_of_physics}
\end{spacing}

\end{document}